\documentclass[12pt]{article}
\pdfoutput=1
\usepackage{epsfig}
\usepackage{amsfonts}
\usepackage{amssymb}

\begin{document}
\newcommand{\nc}{\newcommand}
\nc{\beq}{\begin{equation}} \nc{\eeq}{\end{equation}}
\nc{\beqa}{\begin{eqnarray}} \nc{\eeqa}{\end{eqnarray}}
\nc{\eps}{{\epsilon}}
\nc{\R}{{\cal R}}
\nc{\A}{{\cal A}}
\nc{\K}{{\cal K}}
\nc{\B}{{\cal B}}
\begin{center}

{\bf \Large  Summation of all-loop UV Divergences in Maximally\\[0.4cm] Supersymmetric Gauge Theories } \vspace{1.0cm}

{\bf \large A. T. Borlakov$^{1,3}$, D. I. Kazakov$^{1,2,3}$, D.M. Tolkachev$^{1,4}$\\[0.3cm] and D. E. Vlasenko$^{5}$ }\vspace{0.5cm}

{\it
$^1$Bogoliubov Laboratory of Theoretical Physics, Joint
Institute for Nuclear Research, Dubna, Russia.\\
$^2$Alikhanov Institute for Theoretical and Experimental Physics, Moscow, Russia\\
$^3$Moscow Institute of Physics and Technology, Dolgoprudny, Russia\\
$^4$Stepanov Institute of Physics, Minsk, Belarus\\
$^5$Department of Physics, South Federal State University, Rostov-Don, Russia}\\
E-mail: borlarth@gmail.com, kazakovd@theor.jinr.ru, den3.1415@gmail.com, vlasenko91@list.ru
\vspace{0.5cm}

\abstract{We consider the leading and subleading UV divergences for the four-point on-shell scattering amplitudes in D=6,8,10 supersymmetric Yang-Mills theories in the planar limit.  These theories belong to the class of maximally supersymmetric gauge theories  and presumably possess distinguished properties beyond perturbation theory. In the previous works, we obtained the recursive relations  that allow one to get the leading and subleading divergences in all loops in a pure algebraic way. The all loop summation of the leading divergences  is performed with the help of the differential equations which are the generalization of the RG equations for non-renormalizable theories. Here we mainly focus on solving and analyzing these equations. We discuss the properties of the obtained solutions and interpretation of the results. The key issue is that the summation of infinite series for the leading and the subleading divergences does improve the situation and does not allow one to remove the regularization and obtain the finite answer. This means that despite numerous cancellations of divergent diagrams these theories remain non-renormalizable.}
\end{center}

\section{Introduction}

In recent years maximally supersymmetric gauge theories attracted much attention and served as a theoretical playground promising new insight into the nature of gauge theories beyond usual perturbation theory.
This became possible  due to the development of new computational techniques such as the spinor helicity  and the on-shell momentum superspace formalism~\cite{Reviews_methods}. The most successful examples are the $\mathcal{N}=4$ SYM theory   in $D=4$ ~\cite{BDS4point3loop_et_all}   and the $\mathcal{N}=8$ SUGRA~\cite{N=8SUGRA finiteness}.  These theories are believed to possess several remarkable properties, among which are total or partial cancelation  of UV divergences, factorization of higher loop corrections and possible integrability. The success of factorization leading to the BDS
ansatz~\cite{BDS4point3loop_et_all} for the amplitudes in $D=4$ $\mathcal{N}=4$ SYM stimulated similar activity in other models and dimensions~\cite{Reviews_Ampl_General}.
The universality of the developed methods allows one  to apply them to  SYM theories in dimensions higher than 4 \cite{GeneralDimentions,SpinorHelisity_extraDimentions}.

In this paper,  we focus on the on-shell 4-point amplitude as the simplest structure and  analyze the UV divergences in maximally  SYM theories in D=6,8,10 dimensions in all loops.
For $D>4$ the on-shell amplitudes are IR finite and the only divergences are the UV ones.  Since the gauge coupling $g^2$ in D-dimensions has dimension $[4-D]$,  all these theories are non-renormalizable.

\begin{figure}[htb]
\begin{tabular}{c}
$\frac{\mathcal{A}_4}{\mathcal{A}_4^{(0)}}=1+\sum\limits_L M^{(L)}_4(s,t)=$  \\
\includegraphics[scale=0.35]{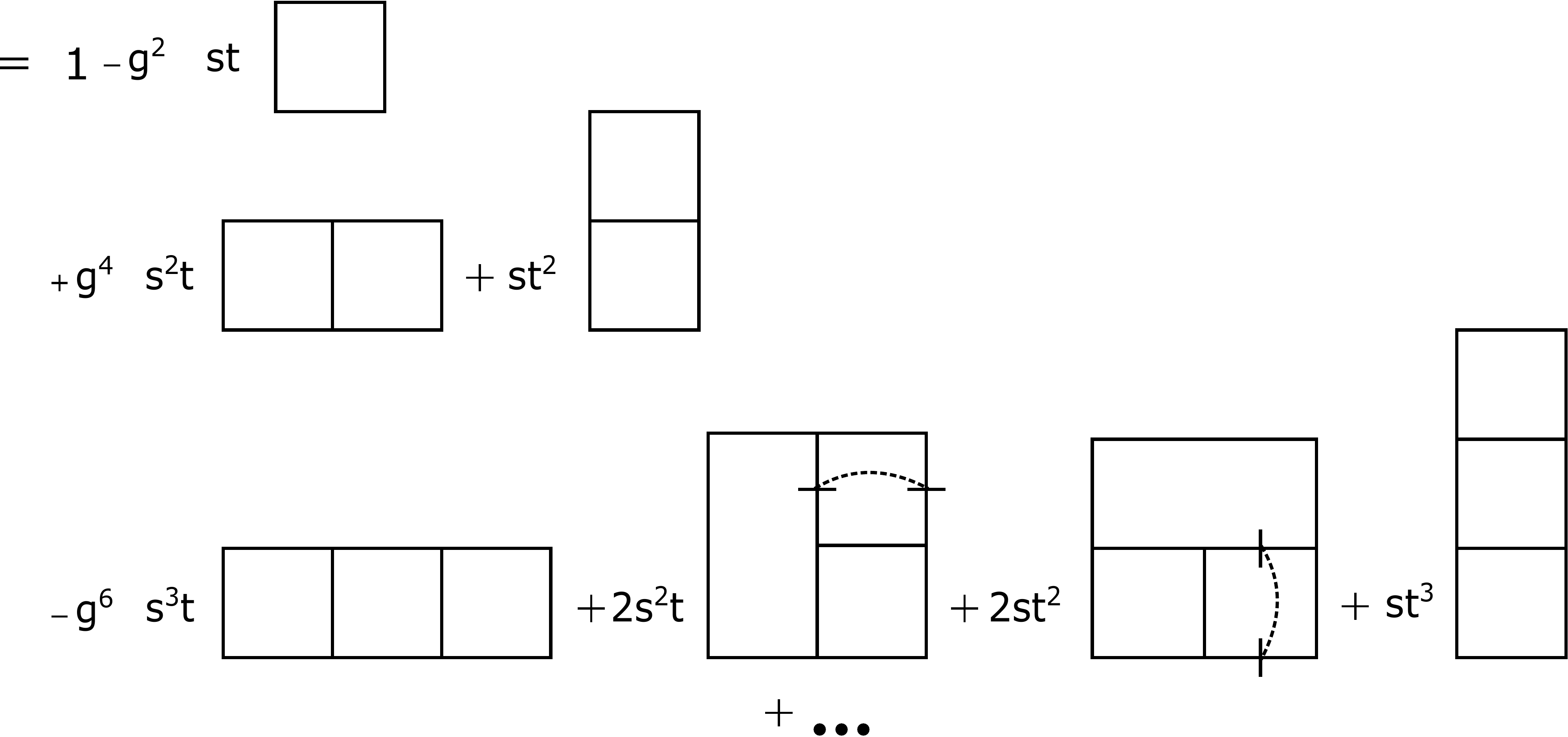} \end{tabular}
\caption{The universal expansion for the four-point scattering amplitude in SYM theories in terms of master integrals.
The connected strokes on the lines mean  the square of the flowing momentum.}\label{expan}
\end{figure}

Applying first the color decomposition of the amplitudes, we are left with the partial amplitudes. Within the spinor-helicity formalism the tree level partial amplitudes  depend on the Mandelstam variables s,t and u and have a relatively simple universal form. The advantage of the superspace formalism is that the tree level amplitudes always factorize  so that the ratio of the loop corrections to the tree level amplitude can be expressed in terms of pure scalar master integrals shown in Fig.\ref{expan}~\cite{Bern:2005iz}.

 Within the dimensional regularization (dimensional reduction) the UV divergences manifest themselves as the pole terms with the numerators being the polynomials over the kinematic variables.
In D-dimensions the first UV divergences start from L=6/(D-4)  loops. Consequently, in D=6 they start from 3 loops. The one loop case is exceptional and in D=8 and D=10 they start already at one loop. Notice that all simple loops as well as triangles completely cancel in all loops. This is the consequence  of maximal supersymmetry and it seems this is maximal it can do. In D=4 this leads to the the cancellation of all the UV divergences since boxes are finite, however, in higher dimensions the UV divergences remain being non-renormalizable by power counting.

In  recent papers~\cite{we1,we2,we3}, we considered the leading and subleading UV divergences of the on-shell scattering amplitudes for all three cases of maximally supersymmetric SYM theories, D=6 (N=2 SUSY), D=8 (N=1 SUSY) and D=10 (N=1 SUSY). We obtained the recursive relations  that allow one to get the leading and subleading divergences in all loops in a pure algebraic way. Then we constructed the differential equations which are the generalization of the RG equations for non-renormalizable theories. Similar to the renormalizable theories, these equations lead to summation of the leading (and subleading) divergences in all loops. Here we concentrate on solving these equations.

It is worth mentioning that PT series in QFT are asymptotic. However, this is true for the full set of diagrams and is not the case of the series for the leading, the subleading, etc. divergences. Indeed, in renoralizable theories the leading divergences simply form the geometric progression as it follows from the one-loop RG equation. It is the beta-function which is given by the asymptotic series. Take same is true for any theory independently on whether it is renormalizable or not.

\section{The Leading Poles in All Loops}
We start with the leading poles and calculate them in all loops. This is possible even in the non-renormalizable case due to the structure of the UV divergences, which follows from the $\R'$-operation. Indeed, according to general theorems, the UV divergences in any loop order after subtraction of divergent subgraphs  are local in the coordinate space~\cite{Bogolubov,Zavialov}.
In \cite{we1,we2}, we exploited this property and obtained the recursion relations that allow one to calculate the leading poles in dimensional regularization algebraically starting with the first (one-loop) ones.
Denoting by  $S_n(s,t)$ and  $T_n(s,t)$  the sum of all contributions  in the  $n$-th order of PT in $s$ and  $t$ channels, respectively, we
got the following recursive relations:\vspace{0.1cm}

{\it D=8, N=1 SYM}
\beqa
&&nS_n(s,t)=-2 s^2 \int_0^1 dx \int_0^x dy\  y(1-x) \ (S_{n-1}(s,t')+T_{n-1}(s,t'))|_{t'=tx+uy}\label{req8}\nonumber \\ &+&
s^4 \int_0^1\! dx \ x^2(1-x)^2 \sum_{k=1}^{n-2}  \sum_{p=0}^{2k-2} \frac{1}{p!(p+2)!} \
 \frac{d^p}{dt'^p}(S_{k}(s,t')+T_{k}(s,t')) \times \nonumber \\
&&\hspace{2cm}\times  \frac{d^p}{dt'^p}(S_{n-1-k}(s,t')+T_{n-1-k}(s,t'))|_{t'=-sx} \ (tsx(1-x))^p, \label{req8}
\eeqa
where  $S_1= \frac{1}{12},\ T_1=\frac{1}{12}$, $u=-s-t$.
The same relation holds for $T(s,t)$ with the replacement $s \leftrightarrow t$ and $T(s,t)=S(t,s)$.\vspace{0.1cm}

{\it D=10, N=1 SYM}
\beqa
&&nS_n(s,t)=-s^3 \int_0^1 dx \int_0^x dy\  y^2(1-x)^2 \ (S_{n-1}(s,t')+T_{n-1}(s,t'))|_{t'=tx+uy}\label{req10}\nonumber \\ &+&
s^5 \int_0^1\! dx \ x^3(1-x)^3 \sum_{k=1}^{n-2}  \sum_{p=0}^{2k-2} \frac{1}{p!(p+3)!} \
 \frac{d^p}{dt'^p}(S_{k}(s,t')+T_{k}(s,t')) \times \nonumber \\
&&\hspace{2cm}\times  \frac{d^p}{dt'^p}(S_{n-1-k}(s,t')+T_{n-1-k}(s,t'))|_{t'=-sx} \ (tsx(1-x))^p,\label{req10}
\eeqa
where $S_1= \frac{s}{5!},\ T_1=\frac{t}{5!}$.\vspace{0.1cm}

{\it D=6 N=2 SYM}

In the case of  $D=6$, since the box diagram is convergent, the  recursive relation has no nonlinear terms and looks like
\beq
nS_n(s,t)=-2 s \int_0^1 dx \int_0^x dy  \ (S_{n-1}(s,t')+T_{n-1}(s,t'))|_{t'=tx+uy}, \ \ \ \     n\geq 4\label{req6}
\eeq
$S_3=-s/3,\ T_3=-t/3$.

The procedure is based on the consistent application of the $\R'$-operation and integration over the remaining triangle and bubble diagrams with the help of Feynman parameters.

These relations take into account all the diagrams of a given order of PT and allow one not only to calculate the leading poles taking the one-loop one as input but to sum all orders of PT. This can be achieved by multiplying both sides of eqs.(\ref{req8})-(\ref{req6}) by $(-z)^{n-1}$, where $z=\frac{g^2}{\epsilon}$ and summing up from n=2 to infinity. Denoting the sum by $\Sigma(s,t,z)=\sum_{n=1}^\infty S_n(s,t) (-z)^n$, we finally get the following differential equations in D=6, 8 and 10 dimensions, respectively,\vspace{0.1cm}

D=6
\beq
\frac{d}{dz}\Sigma(s,t,z)=s-\frac{2}{z}\Sigma(s,t,z)+2s \int_0^1 dx\int_0^x dy (\Sigma(s,t',z)+\Sigma(t',s,z))|_{t'=tx+uy},\label{eq6}
\eeq

D=8
\beqa
&&\frac{d}{dz}\Sigma(s,t,z)=-\frac{1}{12}+2 s^2 \int_0^1 dx \int_0^x dy\  y(1-x)\ (\Sigma(s,t',z)+\Sigma(t',s,z))|_{t'=tx+uy}
\label{eq8}\\
&&-s^4  \int_0^1\! dx \ x^2(1-x)^2 \sum_{p=0}^\infty \frac{1}{p!(p+2)!} (\frac{d^p}{dt'^p}(\Sigma(s,t',z)+\Sigma(t',s,z))|_{t'=-sx})^2 \ (tsx(1-x))^p, \nonumber
\eeqa

D=10
\beqa
&&\frac{d}{dz}\Sigma(s,t,z)=-\frac{s}{5!}+ s^3 \int_0^1 dx \int_0^x dy\  y^2(1-x)^2\ (\Sigma(s,t'z)+\Sigma(t',s,z))|_{t'=tx+uy} \label{eq10}\\
&&-s^5  \int_0^1\! dx \ x^3(1-x)^3 \sum_{p=0}^\infty \frac{1}{p!(p+3)!} (\frac{d^p}{dt'^p}(\Sigma(s,t',z)+\Sigma(t',s,z))|_{t'=-sx})^2 \ (tsx(1-x))^p. \nonumber
\eeqa
The same equations with the replacement $s \leftrightarrow t$ are valid for $\Sigma(t,s,z)$.

Both the recursive relations and the differential equations can be simplified in the case of particular sets of diagrams. For example, for the ladder type diagrams the remained integration over Feynman parameters can be performed explicitly and one is left with the algebraic (for recursive relations) or ordinary differential equations (for the sum of diagrams), which can be explicitly solved. We will consider these solutions in the next section.

\section{Solution of the Equations}
\subsection{The Ladder Case}
Since eqs.(\ref{eq6},\ref{eq8}) and (\ref{eq10}) are integro-differential, their analytical solution is problematic. Therefore, we consider first the case of the ladder type diagrams, which is much simpler and allows for the explicit solution.
As it will be clear later, the ladder type diagrams give the main contribution to the total PT series and may serve as a model for the full answer.\vspace{0.1cm}

{\it D=6}

\begin{figure}[htb!]
\begin{center}
\leavevmode
\includegraphics[width=0.8\textwidth]{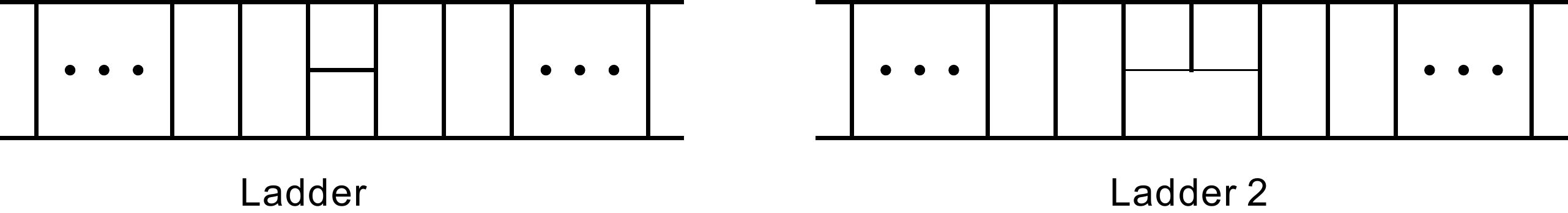}
\end{center}
\caption{The ladder type diagrams in D=6}
\label{laddiag}
\end{figure}

In this case, since the boxes are finite, the s-ladder type diagram of interest contains one tennis-court subdiagram and the ladder of boxes added from the left or right (see Fig.\ref{laddiag} left).

The equation for the ladder diagrams can be obtained from the recursive relations (see \cite{we2}), however, one can also derive it from eq.(\ref{eq6}). Since these diagrams depend only on $s$, the integrals in eq.(\ref{eq6}) for the first term in the r.h.s. drop.  As for the second term, it corresponds to the t-ladder subdiagrams and does not contribute to the ladder approximation. As a result, one has the ordinary differential equation
\beq
\frac{d \Sigma_L(s,z)}{dz}=s-\frac 2z \Sigma_L(s,z)+s\Sigma_L(s,z), \ \ \ \ \Sigma_L(s,0)=0.
\eeq
Note that $\Sigma_L(s,z)$ is dimensionless and depends on a single dimensionless argument $sz$.

The solution to this equation is
\beq
\Sigma_L(s,z)=\frac{2}{s^2z^2}(e^{sz}-1-sz-\frac{s^2z^2}{2}).\label{lad6}
\eeq
And the same for the vertical ladder with the replacement $s\leftrightarrow t$.

One can see that the obtained solution tends either to infinity or to a constant when $z\to\infty\ (\epsilon\to 0)$
depending on the sign of $s$. We will see that the full solution has the same tendency. We  discuss the consequences of this behaviour below.

 One can get a similar expression  for the next sequence of diagrams shown in Fig.\ref{laddiag} (right). This is also a ladder type diagram starting with the t-channel tennis-court at four loops. In this case, the resulting expression contains two contributions, one proportional to $s$ and the other to $t$ times the functions which depend only on $s$.
For this reason one has two coupled recursive relations and hence two coupled differential equations. The resulting expression has the form
\beqa
\Sigma_{L2}(s,t,z)&=&\frac{1}{2s^2z^2}\left[27(e^{sz/3}-1-\frac{sz}{3}-\frac 12\frac{s^2z^2}{9}-\frac 16\frac{s^3z^3}{27})(1+2\frac ts)\right.\nonumber\\
&&\left.-(e^{sz}-1-sz-\frac 12 s^2z^2-\frac 16 s^3z^3)\right].\label{lad62}
\eeqa
This expression has properties similar to the previous one. Depending on the sign of $s$, it tends either to infinity or to a constant.
We will see later that the sum of two ladders  (\ref{lad6}) and (\ref{lad62}) gives a better approximation to the full answer.\vspace{0.1cm}

{\it D=8}

In the case of D=8, the ladder diagrams start already with one loop. They also depend only on $s$ so that in eq.(\ref{eq8}) all integrals are trivial for the first terms in the bracket while the second terms do not contribute to the s-ladder like in the previous case. Then eq.(\ref{eq8}) is reduced to the ordinary nonlinear differential equation
\beq
\frac{d \Sigma_L(s,z)}{dz}=-\frac{1}{3!}+\frac{2}{4!} \Sigma_L(s,z)-\frac{2}{5!}\Sigma_L^2(s,z), \ \ \ \ \Sigma_L(s,0)=0.
\eeq
Note that $\Sigma_L(s,z)$ here is also dimensionless and depends on a single dimensionless argument $s^2z$.

This is the Riccati type equation with constant coefficients. Its solution has the form
\beq
\Sigma_L(s,z)=-\sqrt{5/3} \frac{4 \tan(zs^2/(8 \sqrt{15}))}{1 - \tan(zs^2/(8 \sqrt{15}))\sqrt{5/3}}.\label{lad8}
\eeq

This function possesses an infinite number of periodical poles and has no limit when $z\to\infty\ (\epsilon\to 0)$ independently of kinematics. We will see that as in the previous case of D=6 the full solution inherits this property.\vspace{0.1cm}

{\it D=10}

The situation in this case reminds that for D=8 but is  more complicated since the genuine box diagram in D=10 contrary to D=8 is not a constant but is proportional to $(s+t)$. Consequently, the s-ladder has dimension $m^2$ and consists of two parts, one proportional to s and the other to t times the function of $s$

\beq
\Sigma_L(s,t,z)=s\Sigma_{Ls}(s,z)+t\Sigma_{Lt}(s,z). \label{d10lad}
\eeq

Similar to the D=8 case, eq.(\ref{eq10}) is reduced to the ordinary nonlinear differential equation, however, here we obtain two coupled equations for  $\Sigma_{Ls}(s,z)$ and $\Sigma_{Lt}(s,z)$. The simplest way to get them is to use the recursive relations~\cite{we2}
\beqa
\frac{d \Sigma_{Lt}(s,z)}{dz}&=&
-\frac{1}{5!} +\frac{4}{7!}\Sigma_{Lt}(s,z) - \frac{1}{3*7!}\Sigma_{Lt}^2(s,z), \ \ \ \ \ \ \ \ \  \ \ \ \ \ \ \ \ \ \Sigma_{Lt}(s,0)=0,\\
\frac{d \Sigma_{Ls}(s,z)}{dz}&=& - \frac{1}{5!} +\frac{2}{3*5!}\Sigma_{Ls}(s,z) - \frac{12}{7!}\Sigma_{Lt}(s,z)\nonumber \\&-&\frac{3!}{7!}\left(\Sigma_{Ls}^2(s,z)-\Sigma_{Ls}(s,z)\Sigma_{Lt}(s,z)+\frac{5}{18}\Sigma_{Lt}^2(s,z)\right), \   \Sigma_{Ls}(s,0)=0.
\eeqa
Note that both functions are dimensionless and depend on a single dimensionless argument $s^3z$.

The solution to the first equation is
\beq
\Sigma_{Lt}(s,z)=3 \left(2 + \sqrt{10}\tan\left[\frac{-\sqrt{10} zs^3 - 5040 \arctan[\sqrt{2/5}]}{5040}\right]\right),\label{lad10}
\eeq
while for the second one it might be expressed in the form
\beq
\Sigma_{Ls}(s,z) = \frac{1}{2}\Sigma_{Lt}(s,z) + \Delta(s,z), \label{sol10}
\eeq
where the function $\Delta(s,z)$ obeys the nonlinear differential equation
\beq
\frac{d \Delta(s,z)}{dz}=
- \frac{1}{2*5!} +\frac{2}{3*5!}\Delta(s,z) - \frac{6}{7!}\Delta^2(s,z), \ \ \Delta(s,0)=0.\label{dd}
\eeq
This is also a dimensionless function of a single dimensionless variable $s^3z$. The solution to eq.(\ref{dd}) is
\beq
\Delta(s,z)=-\frac{(3 (14 + \sqrt{70}) (-1 + e^{zs^3/(36 \sqrt{70})})}{2 (19 + 2 \sqrt{70} - 9 e^{zs^3/(36 \sqrt{70})})}.\label{ladd}
\eeq
Summarizing one has the following expression for the ladder diagram
\beq
\Sigma_L(s,t,z)=s\left(\frac 12\Sigma_{Lt}(s,z)(1 +2\frac ts)+\Delta(s,z)\right), \label{totlad10}
\eeq
where $\Sigma_{Lt}(s,z)$ and $\Delta(s,z)$ are given by eqs.(\ref{lad10}) and (\ref{ladd}), respectively.
The behaviour of $\Sigma_{Lt}$ is similar to $\Sigma_L$ in the  D=8 case. It possesses an infinite number of periodical poles. The function $\Delta$ has a single pole for positive values of $s$.

\subsection{The General Case}

Here we analyze the full equations (\ref{eq6},\ref{eq8},\ref{eq10}) which reproduce the sum of all the diagrams. As one can see, these equations are integro-differential and  cannot be treated analytically. Instead, we perform a numerical study of these equations though this is also not straightforward. The reason is that one cannot use the standard recursive algorithm since the functions in the r.h.s. stand under the integral sign and depend on the integration variables.

Therefore, we apply the following method
which is the combination of numerical and successive approximation approaches. At the beginning we choose some constant value of the function $\Sigma(s,t,z)=\Sigma_0(s,t,z) = const$ from which the procedure starts. If we consider the interval beginning from $z_0=0$, the obvious choice is $const=0$. Then, we substitute it into the r.h.s.  of the equation and perform the formal integration. Replacing the derivative in the l.h.s.  by the finite difference  $(\Sigma_1(s,t,z)-\Sigma_0(s,t,z))/\Delta z$, we finally get the next approximation for $\Sigma$
\beq
\Sigma_1(s,t,z)=\Sigma_0(s,t,z)+\Delta z * r.h.s,  \eeq
which is now a polynomial over $s$ and $t$. At this step, the r.h.s. is calculated with $\Sigma_0(s,t',z)$ equal to a constant.

Moving forward to the next step, we substitute the obtained polynomial for $\Sigma_1(s,t,z)$ into the r.h.s., change the arguments $t\rightarrow tx+uy$ and $t \rightarrow -sx$, and perform the integration. This generates the next approximation value of $\Sigma$: $\Sigma_2(s,t,z)$. Continuing this way we generate the higher order polynomials of $s$ and $t$ at each step. However, starting from 3-4 iterations the length of polynomials becomes too high to continue.
At this step, we evaluate $\Sigma$ with fixed values of s and t, for instance, s=t=1. The calculated value gives us a constant which we identify with the value of $\Sigma$ at the point $z_0+\Delta z$. We then use it to start the same procedure again for the next point.

This way we calculate the values of $\Sigma$ at the points along the axis $z=z_0+ \Delta z * n$.
Then we interpolate all the obtained points getting a smooth function. Note that when applying numerical computation one has to fulfill the following requirements: a sufficient degree of smoothness of the calculation and minimization of time spent on its implementation. Experimentally, the step $\Delta z = 0.1$ has been found to meet these requirements and also made the solution stable. And though this method is not justified, numerical results show a very good approximation being applied to known functions.

It is worth mentioning that after the evaluation of the function $\Sigma$, we can replace its argument having in mind that on dimensional grounds it depends on dimensionless combinations $zs, zs^2$ and $zs^3$ (and the same for $t$) for $D=6,8$ and $D=10$, respectively. We use these substitutions in the next section to plot the results.

One should also mention that in $D=8,10$ the form of eq.(\ref{eq8},\ref{eq10}) is not very suitable for numerical analysis because the second term contains the infinite sum with an infinite number of derivatives. The necessity to cut this sum makes the numerical solution unstable. To avoid this problem, we notice that
the construction reminds the usual shift operator with slightly changed coefficients. One can prove that the infinite sum might be removed by introducing two additional integrations using the following formula:
\beqa
&&\sum_{p=0}^\infty \frac{(BC)^p k!}{p!(p+k+1)!} \left(\frac{d^p}{dA^p}f(A)\right)^2\\=
&&\frac{1}{2\pi}\int_{-\pi}^{\pi} d\tau \int_0^1 d\xi(1-\xi)^k f(A+exp(i\tau)B\xi) f(A+exp(-i\tau)C). \nonumber
\eeqa
Two additional integrals do not cause any trouble for numerical integration. We use this trick for calculations in the case of $D=8$ and $D=10$.

The realization of the advocated procedure in the case of $D=8$ is presented in the form of the Mathematica code written below.
\begin{footnotesize}
\begin{verbatim}
L = {0}; (*starting value*)
h = 0.1; (*step value*)
Do[l = {L[[d]]};
 For[i = 1, i <= 3, i++,
  l = Append[l,
    l[[i]] + (-h/12 +
       2  h s^2  Integrate[
         Integrate[
          y (1 - x) ( l[[i]] + ( l[[i]] /. {t -> s, s -> t})) /.
           t -> t x - t y - s y, {y, 0, x}], {x, 0, 1}] -
       If[i > 2,
        s^4  h Integrate[
          x^2 (1 - x)^2 Integrate[
            Integrate[(1 - kc) (l[[i]] /.
                t -> -s x +
                  Exp[I ta] t (1 - x) kc) ((l[[i]] /. {t -> s,
                   s -> t}) /. t -> -s x + Exp[-I ta] s  x) , {kc, 0,
              1}], {ta, -Pi, Pi}], {x, 0, 1}], 0])]];
 z = (l + (l /. {t -> s, s -> t}))/2;
 L = Append[L,
   Delete[z, {{1}, {2}, {3}}][[1]] /. s -> 1 /. t -> 1], {d, 3}]
\end{verbatim}
\end{footnotesize}

A drawback of this approach is hidden in the fact that the procedure does not depend on z, i.e., we reconstruct the form of the solution but do not fix it on the z-axis. In other words, the position of the solution is not absolute but relative. In the  region starting from $z=0$ to the first pole, this problem is absent since we know that at the beginning the function equals 0. In the next region between the first and the second poles, one should choose the starting point which is close to the pole and start the procedure until it reaches the second pole. Then one continues the same way for the next intervals.

We performed the described calculations  for all three cases $D=6,8,10$. The results are presented in the next section.

\section{Comparison of PT, Pade, Ladder and Numerics}
In this section, we compare the results of calculation of the leading order divergences
given by eqs.(\ref{eq6},\ref{eq8},\ref{eq10}) using the PT, Pade approximation, Ladder approximation and numerical solution described above.

For comparison we use the first 15 terms of PT generated with the help of recursive relations. This seems to be by far enough since the successive terms of PT fall rapidly.

The next step is the use of the Pade approximation. It is not always stable since the Pade approximants sometimes possess fictitious poles. This is a well known feature and we tried to avoid it taking mainly diagonal approximants.
With 15 terms of PT the [6/6], [6/7] and [7/7] approximants are almost identical and give a smooth function.

The third  curve on the plots corresponds to the ladder approximation. Here the analytical solutions are given by eqs.(\ref{lad6},\ref{lad8},\ref{lad10},\ref{ladd}) from the previous section.  In the case of D=6 we also considered the second ladder which is based on the tennis-court diagram in the t-channel (see Fig.\ref{laddiag}) and is given by eq.(\ref{lad62}).

At last, we plot the numerical solution obtained via the iteration procedure described above. In the case when the function possesses poles we built the numerical solution separately for each finite interval.

The function of interest $\Sigma(s,t,z)$ is the function of three variables. However, as it was already mentioned, on dimensional grounds it has only two independent dimensionless arguments. In $D=6,8$ and $10$ dimensions they are  $zs, zt$, $zs^2, zt^2$ and $zs^3,zt^3$, respectively. For a better presentation we constructed both the two-dimensional plots with $t=s$ and the 3-dimensional ones in the $s-t$ plane.\vspace{0.1cm}

{\it D=6}

In D=6 the PT series looks like
\beqa
\Sigma_{PT}(s,t,z)&=& \frac{(s + t)z}{3} + \frac{(s^2 + st + t^2)z^2}{18} + \frac{(5s^3 + 2s^2t + 2st^2 + 5t^3)z^3}{540}
\nonumber\\
&+&\frac{ (25 s^4 + 8 s^3 t - 2 s^2 t^2 + 8 s t^3 + 25 t^4) z^4}{19440}  + ... ,\label{pt6}
\eeqa
where the dots stand for the higher order terms. We used 15 terms for numerical comparison with the other approaches.

From eq.(\ref{pt6}) we constructed the diagonal Pade approximant [7/7]  as a function of a new variable $x=zs$ in the case when $t=s$. It has the form
\beqa
\Sigma_{Pade}(x)= && \frac{0.67 x + 0.067 x^2 + 0.0010 x^3 +  0.00014 x^4 + 4.6\cdot10^{-5} x^5 +}{1 - 0.15 x + 0.00013 x^2  +
 0.0011 x^3 - 4.5\cdot10^{-5} x^4 - 2.1\cdot10^{-6} x^5 +} \rightarrow  \nonumber \\
&& \leftarrow \frac{ + 3.7\cdot10^{-6} x^6 + 1.2\cdot10^{-7} x^7}{+ 1.7\cdot10^{-7} x^6 -2.1\cdot10^{-9} x^7}.
 \label{pade6}
\eeqa

The ladder approximation is given by eq.(\ref{lad6}) and the second ladder by eq.(\ref{lad62}) with $x=zs$.
The numerical solution starts from $z=0$ and has only one interval in this case.

\begin{figure}[htb!]
\begin{center}
\leavevmode
\includegraphics[width=0.46\textwidth]{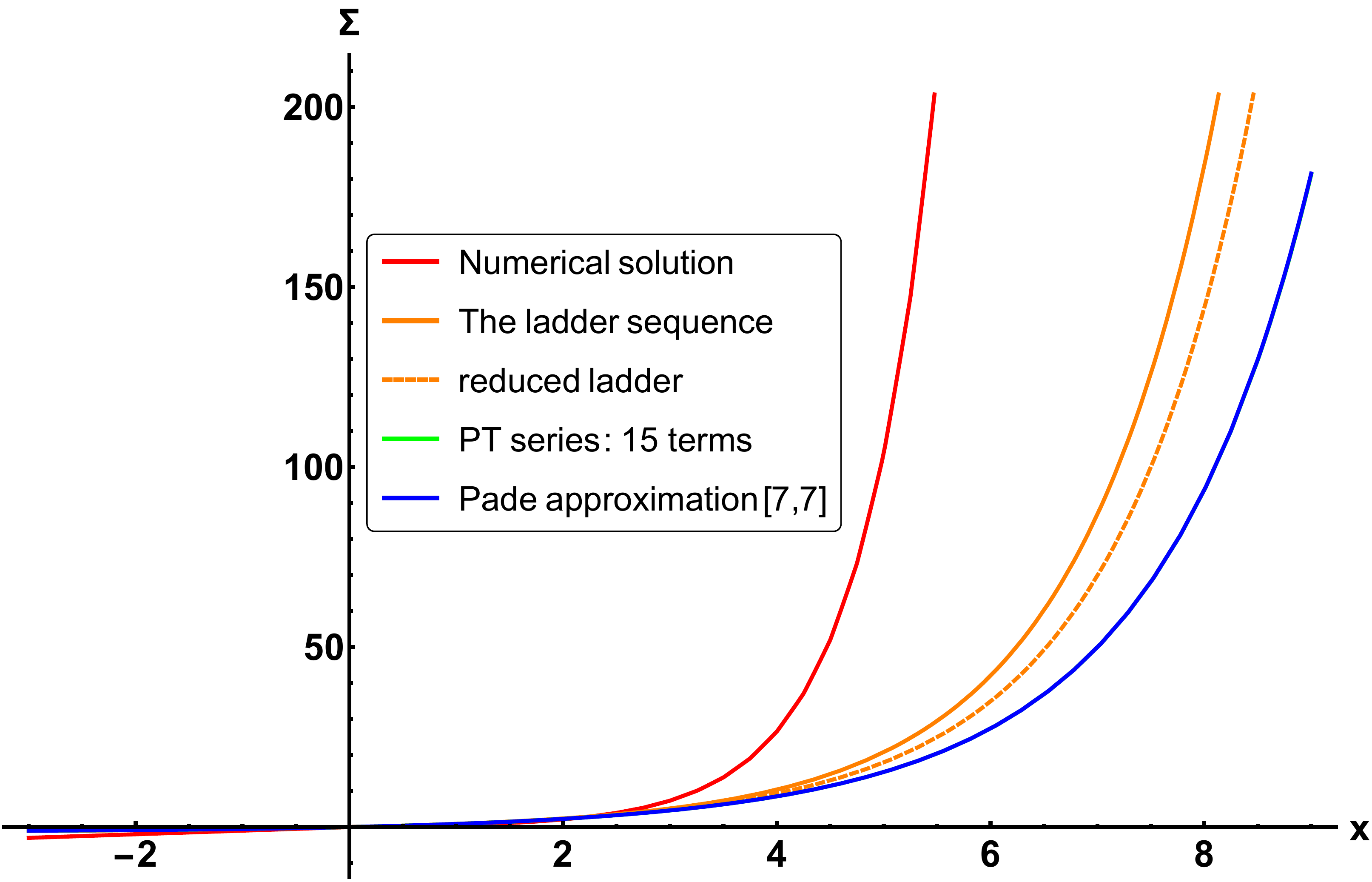}
\end{center}
\caption{Comparison of various approaches to solve eq.(\ref{eq6}): PT, Pade, Ladder and Numerics. The PT curves and Pade coincide in a given interval.
The red line is the numerical solution, the green one is the PT, the blue one is the Pade approximation. The yellow line represents the ladder analytical solution}
\label{allloop6}
\end{figure}

To demonstrate the behaviour of the function $\Sigma$ obtained by various approaches and to compare them all, we plot them together . The first plot
shown in Fig.\ref{allloop6} contains five different curves
corresponding to five approaches: the PT, Pade approximation, ladder approximation, two-ladder approximation and the numerical solution.

\begin{figure}[htb!]
\begin{center}\vspace{0.5cm}
\leavevmode
\includegraphics[width=0.32\textwidth]{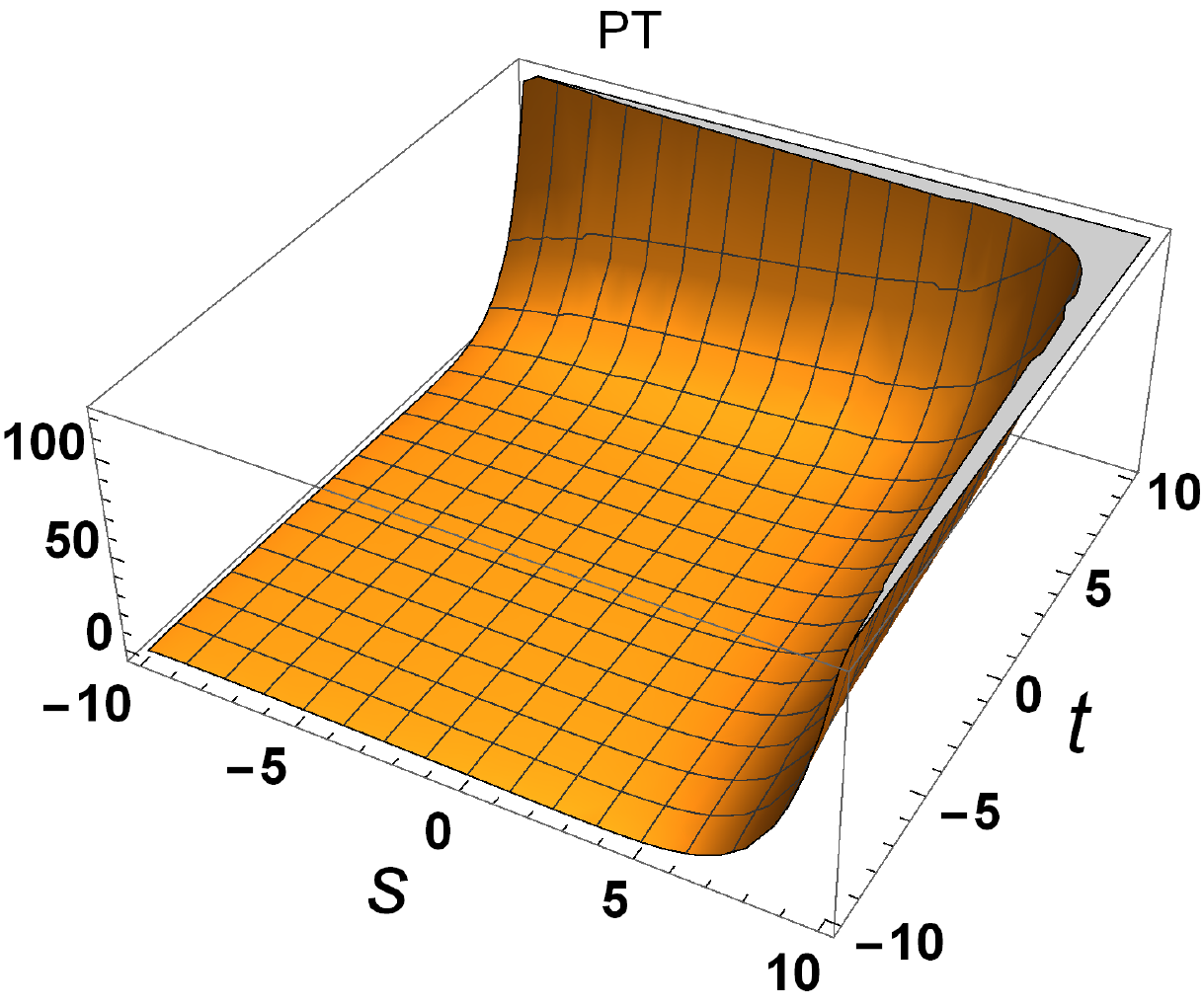}
\includegraphics[width=0.32\textwidth]{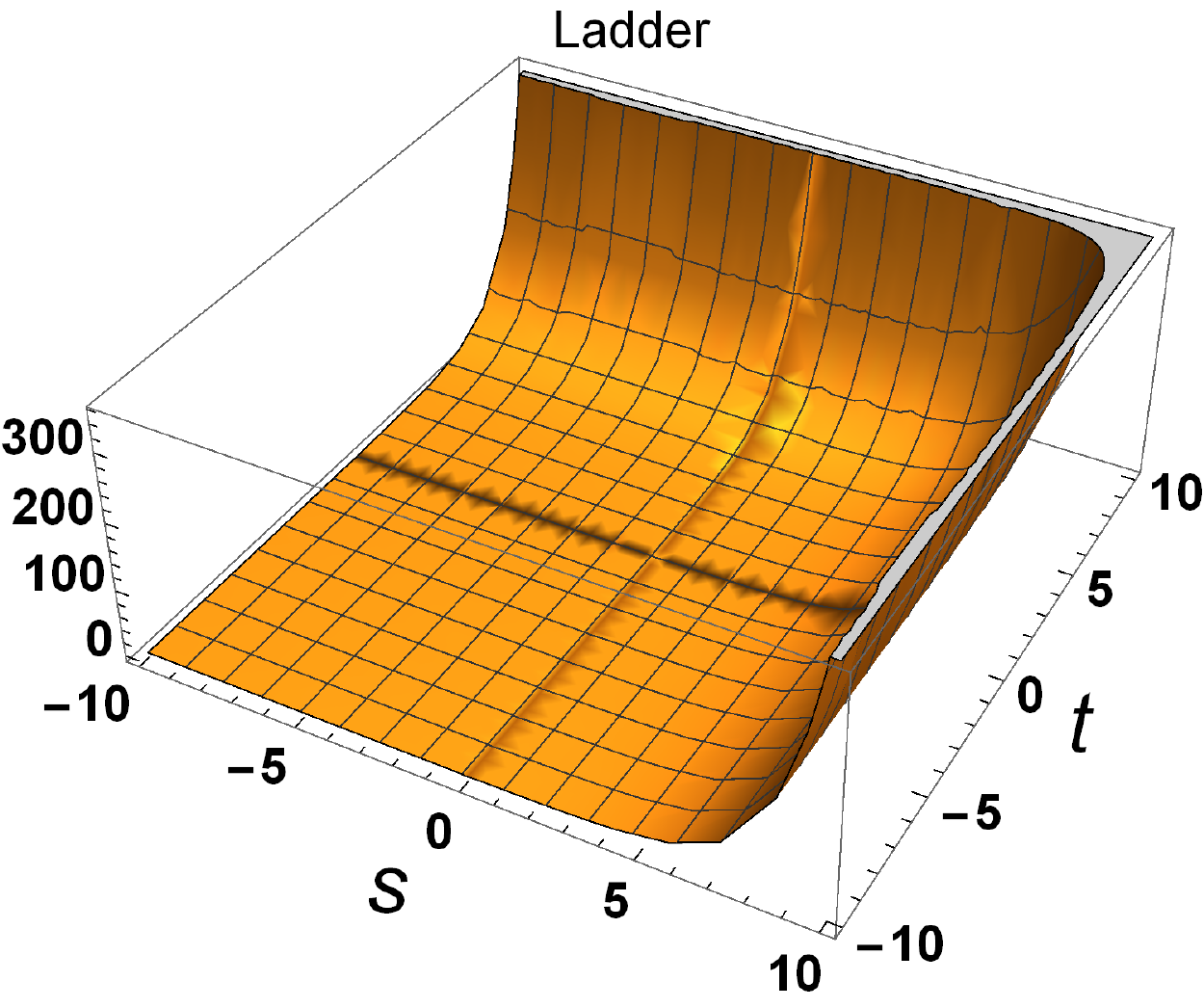}
\includegraphics[width=0.32\textwidth]{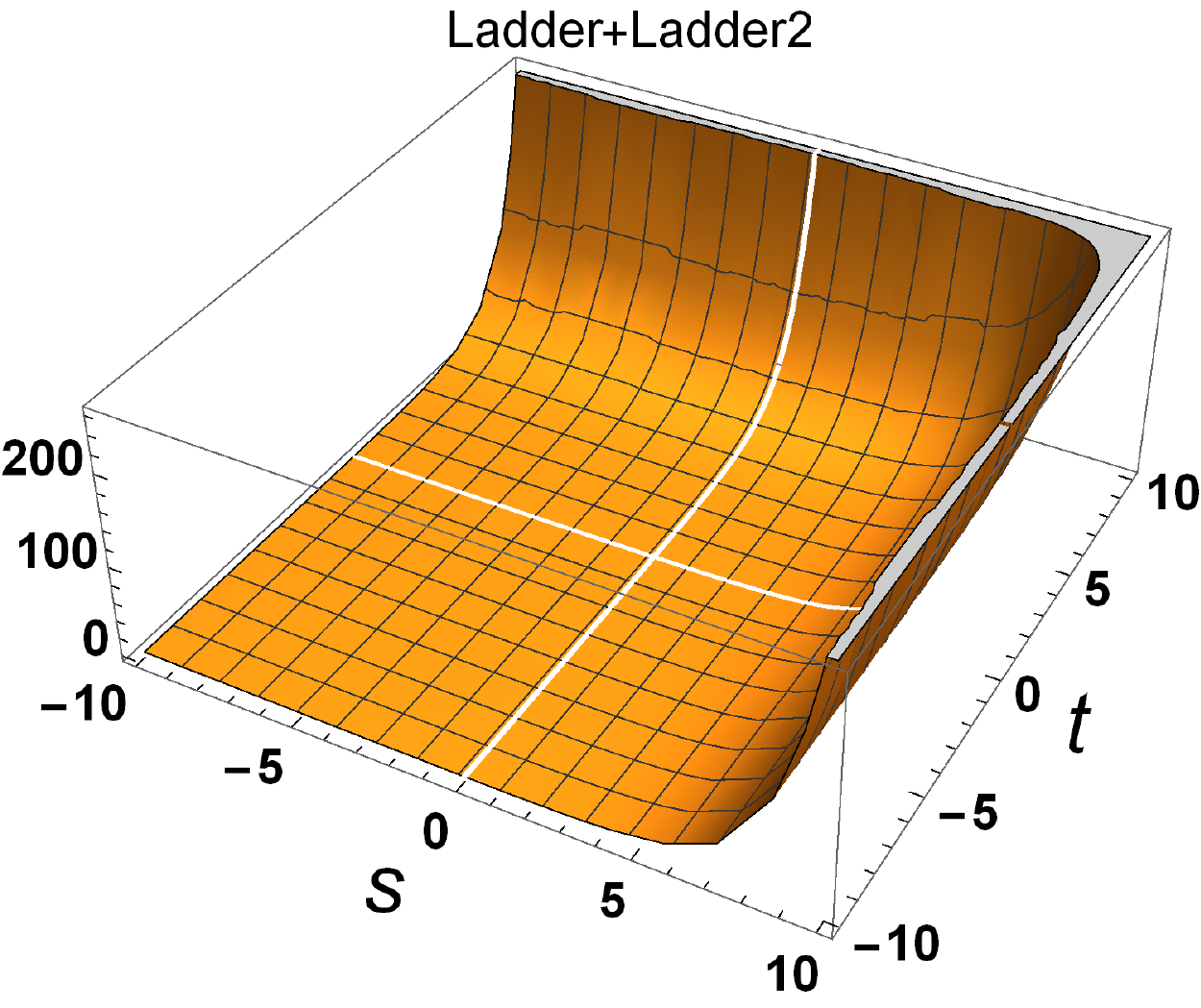}
\end{center}
\caption{Comparison of  PT, Ladder and Two-Ladder approximations}
\label{3d6}
\end{figure}

The second plot is the 3-dimensional one shown in Fig.\ref{3d6}. Here we plot PT, the ladder approximation and the two-ladder approximation.

One can see on the first plot that all the curves practically have the same behaviour.  Analytically, it is perfectly described by the ladder approximation. This is also confirmed by the 3-dimensional plot shown in Fig.\ref{3d6}. The inclusion of the second ladder does not change the solution qualitatively but provides a better correspondence with PT. The function $\Sigma$ has no limit when $x\to\infty$ ($\epsilon\to0$) for $s>0$ and tends to a fixed point when $s<0$.  This limit would correspond to removing the UV regularization. One can see that summation of the whole infinite series does not improve the situation. One can not just remove the UV regularization and get a finite theory.\vspace{0.1cm}

{\it D=8}

In the case of D=8, the PT series starts already from one loop and has the form
\beqa
\Sigma_{PT}(s,t,z)&=& \frac{z}{6}+\frac{s^2 + t^2}{144}z^2 +\frac{15 s^4 - s^3 t + s^2 t^2 - s t^3 + 15 t^4}{38880} z^3\label{pt8}    \\
&+& \frac{8385 s^6 - 268 s^5 t + 206 s^4 t^2 - 192 s^3 t^3 + 206 s^2 t^4 -
   268 s t^5 +8385 t^6}{391910400}z^4  + ...      \nonumber
\eeqa
For $t=s$ the [7/6] the Pade  approximant is
\beqa
\Sigma_{Pade}(x)=&& \frac{1}{s^2}\frac{0.17 x - 0.017 x^2 + 0.00040 x^3 + 0.000014 x^4 - 7.1\cdot10^{-7} x^5 +}{1 - 0.19 x + 0.014 x^2 -
 0.00046 x^3 + 6.9\cdot10^{-6} x^4 - 1.5\cdot10^{-8} x^5 -} \rightarrow  \nonumber \\
&& \leftarrow \frac{ +
 7.5\cdot10^{-9} x^6 + 1.2\cdot10^{-10} x^7}{ - 5.5\cdot10^{-10} x^6},
 \label{pade8}
\eeqa
where now  $x=zs^2$.

The ladder approximation is given by eq.(\ref{lad8}).
The numerical solution starts from $z=0$ and goes up to the first pole $z=z_1$. Then we start it again at some $z>z_1$ and arrive at the second pole at $z=z_2$, etc. The poles coincide with those of the ladder approximation (\ref{lad8}) thus confirming that the ladder approximation gives a  correct picture of the full answer.

The comparison of various curves for $t=s$ is shown in Fig.\ref{allloop8}.
\begin{figure}[htb!]
\begin{center}
\leavevmode
\includegraphics[width=0.45\textwidth]{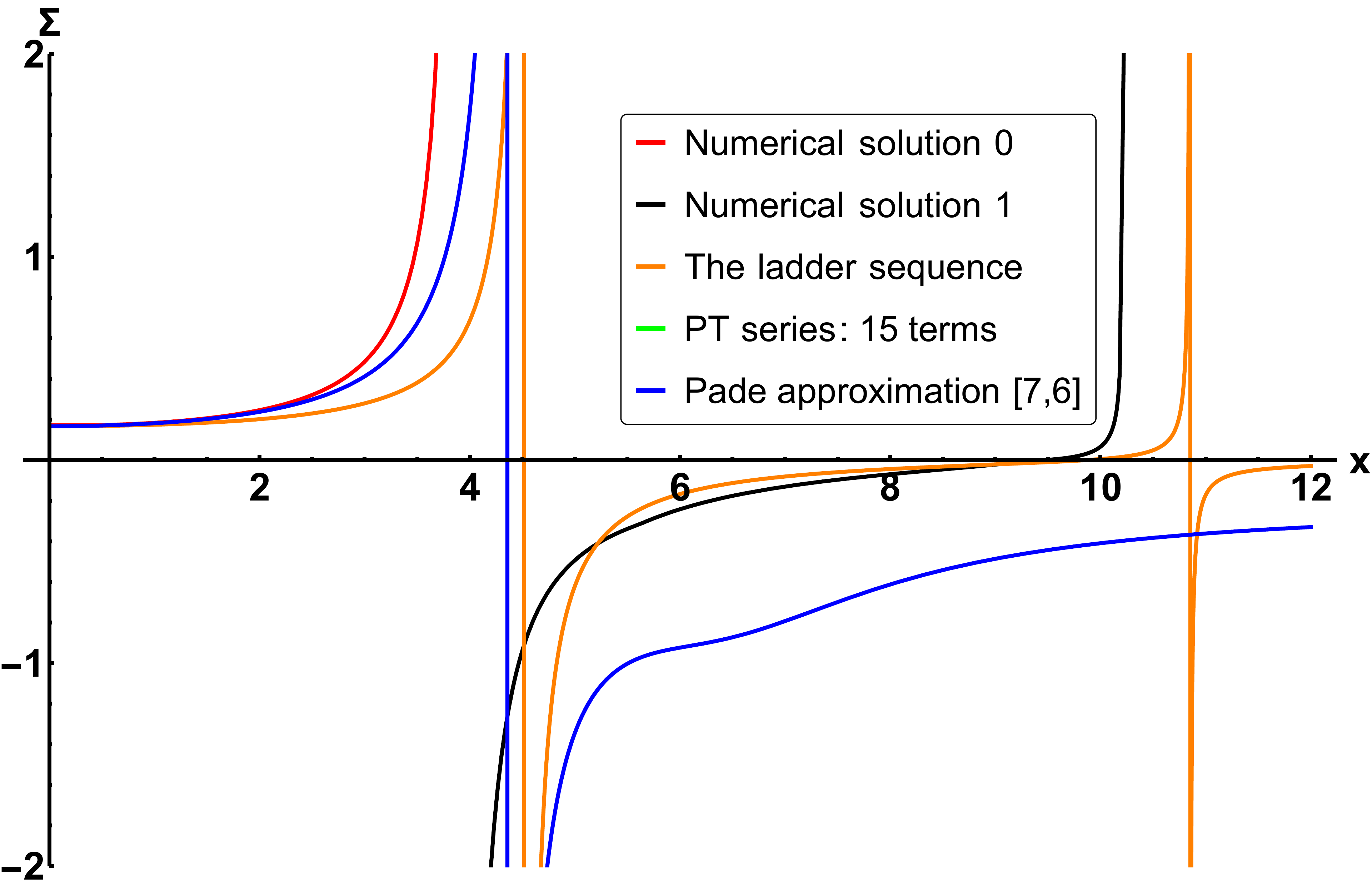}
\label{allloop8}
\end{center}
\caption{Comparison of various approaches to solve eq.(\ref{eq8}) . The red and black lines are the numerical solutions The green one is the PT. The blue one is the Pade approximation.
The yellow one
represents the Ladder analytical solution}
\label{allloop8}
\end{figure}
\begin{figure}[htb!]
\begin{center}
\leavevmode
\includegraphics[width=0.3\textwidth]{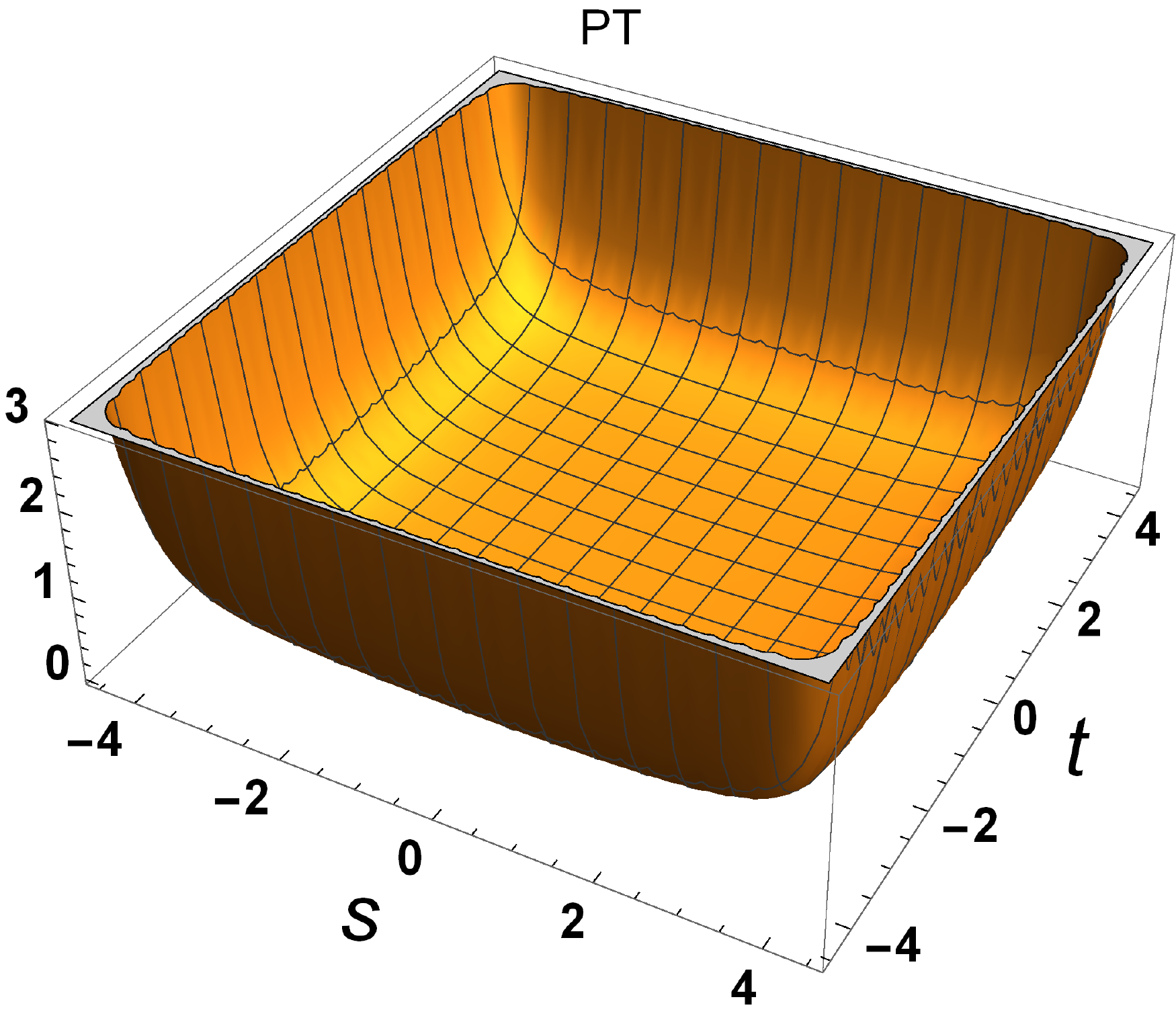}
\includegraphics[width=0.3\textwidth]{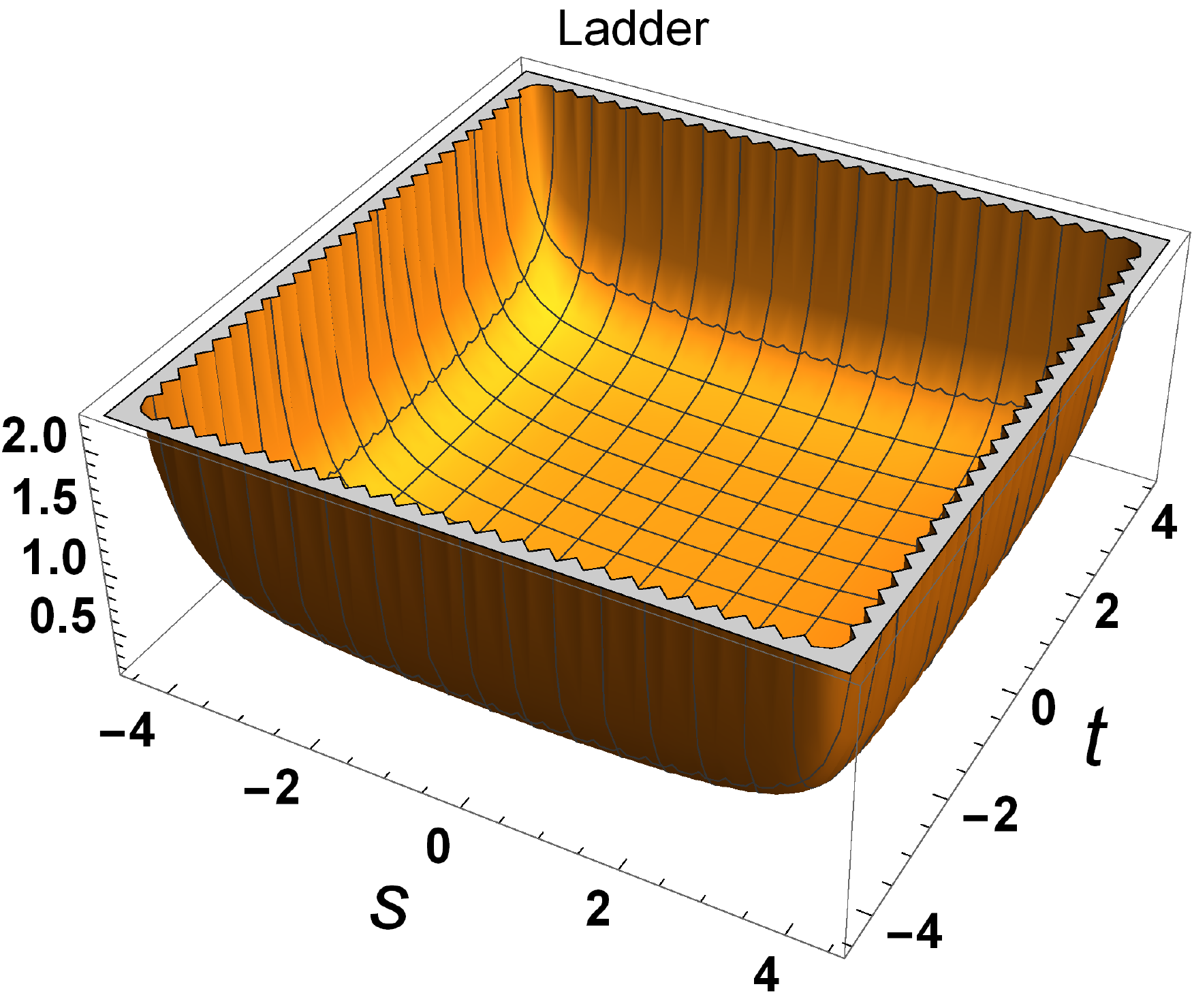}
\includegraphics[width=0.3\textwidth]{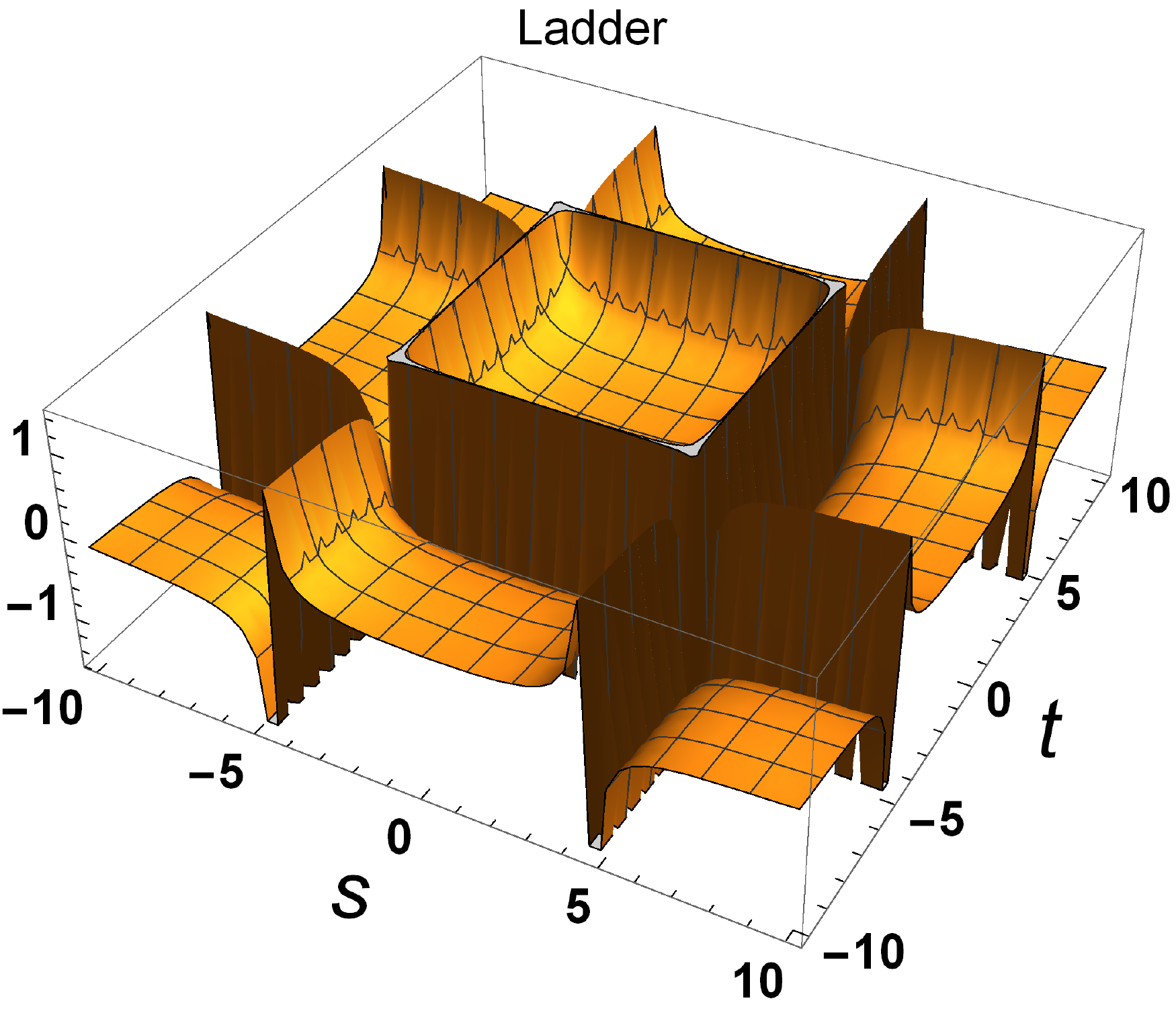}
\end{center}
\hspace{3cm} a \hspace{4cm} b \hspace{5cm} c
\caption{Comparison of PT(a)  and the  ladder approximation (b) in the region up to the first pole. The last plot (c) shows the ladder approximation beyond the first pole. One can clearly see the pole structure of the function $\Sigma$. }
\label{3d8}
\end{figure}
One can see that in the first interval all curves practically coincide. The PT curve exists only in the interval below the first pole. The Pade curve reproduces the first pole but fails with the other ones. The numerical curve reproduces both poles and  is close to the ladder approximation.

We present also the 3-dimensional plot in the $s-t$ plane for $z=1$ in Fig.\ref{3d8}.
One can see that the ladder diagrams, like in the case of $D=6$,  give a very accurate  approximation to a full PT and allow one to go beyond the first pole. The comparison of the ladder approximation and numerical solution of the full equation justifies our conclusion that the ladder approximation reproduces the correct behaviour of the function.

Again, we have to admit that the limit $x\to\infty$ ($\epsilon\to 0$) does not exist. The function $\Sigma$ has an infinite number of periodic poles  for any choice of kinematics. Therefore, the UV finiteness is not reached when the sum over all loops is taken into account.\vspace{0.1cm}

{\it D=10}

This case is quite similar to the $D=8$ one. Now the PT series now is
\beqa
&&\Sigma_{PT}(s,t,z)  = \frac{(s + t)z}{120} + \frac{(4 s^4 + s^3 t + s t^3 + 4 t^4)z^2}{302400} + \\ \label{pt10}
&&+\frac{(2095 s^7 + 115 s^6 t + 33 s^5 t^2 - 11 s^4 t^3 - 11 s^3 t^4 +
  33 s^2 t^5 + 115 s t^6 + 2095 t^7)z^3}{68584320000} +  ...  \nonumber
\eeqa
while the [6/7] Pade approximation for $t=s$ reads
\beqa
\Sigma_{Pade}(x)=&& \frac{1}{s^2}\frac{0.017 x + 0.00025 x^2 + 6.5\cdot10^{-7} x^3 - 5.7\cdot10^{-10} x^4 -
 }{1 + 0.013 x +
 9.4\cdot10^{-6} x^2 - 1.1\cdot10^{-7} x^3 - 7.2\cdot10^{-11} x^4 +} \rightarrow  \nonumber \\
&& \leftarrow \frac{  - 2.1\cdot10^{-12} x^5 +  2.6\cdot10^{-16} x^6 + 7.3\cdot10^{-19} x^7}{+ 1.9\cdot10^{-13} x^5 - 6.4\cdot10^{-17} x^6 + 4.6\cdot10^{-21} x^7},
 \label{pade8}
\eeqa
where $x=zs^3$.

The ladder approximation is given by eqs.(\ref{totlad10}) and the numerical one is again constructed first for the interval from $z=0$ to the first pole and then continued to the second one, etc.

\begin{figure}[htb!]
\begin{center}
\leavevmode
\includegraphics[width=0.5\textwidth]{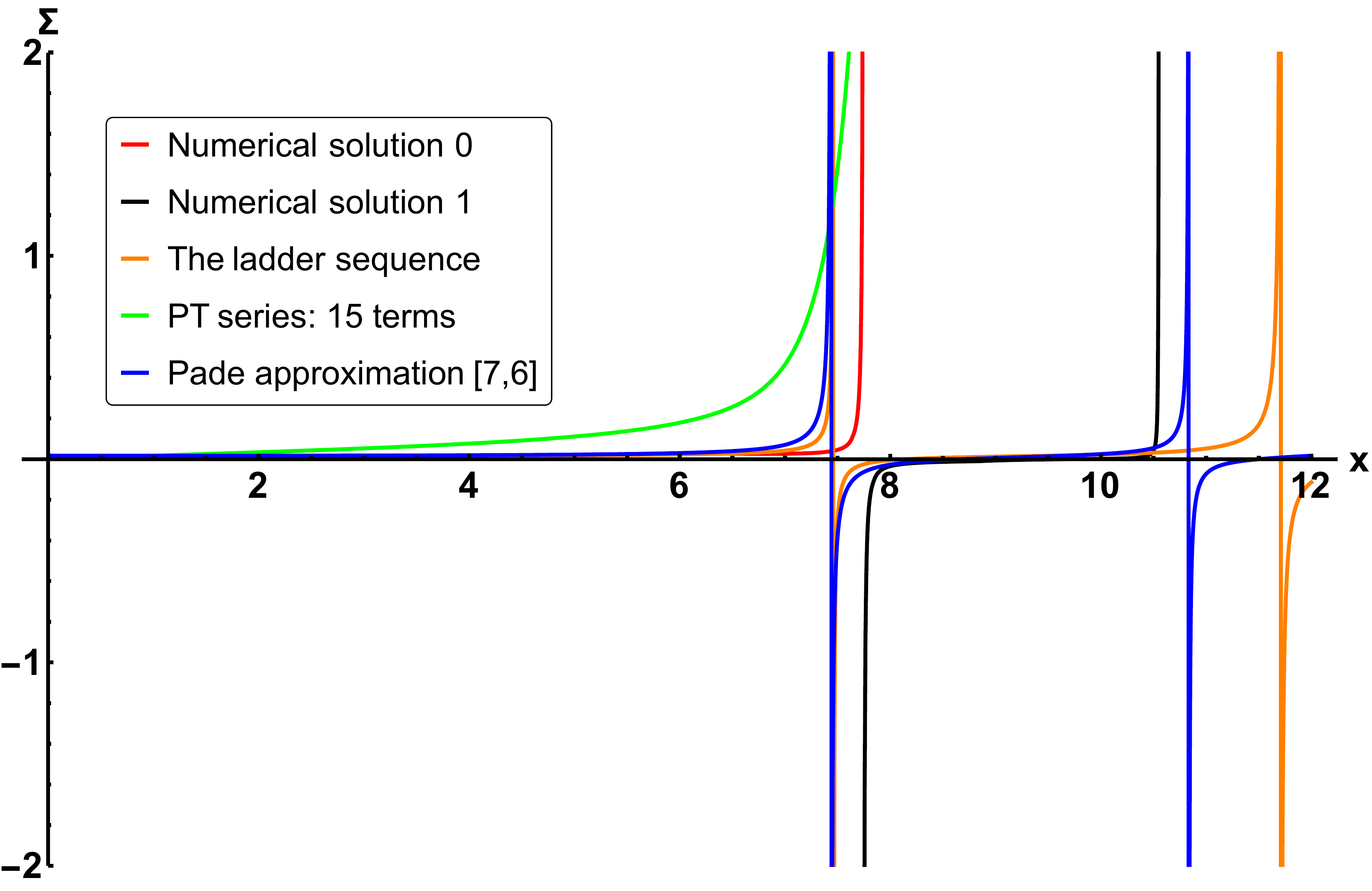}
\end{center}
\caption{Comparison of various approaches to solve eq.(\ref{eq10}) . The red and black lines are the numerical solutions described in the previous section before the first pole and the first and the second ones. The green one is the PT. The blue one is the Pade approximation. The yellow line represents the Ladder analytical solution}
\label{allloop10}
\end{figure}

\begin{figure}[htb!]
\begin{center}
\leavevmode
\includegraphics[width=0.3\textwidth]{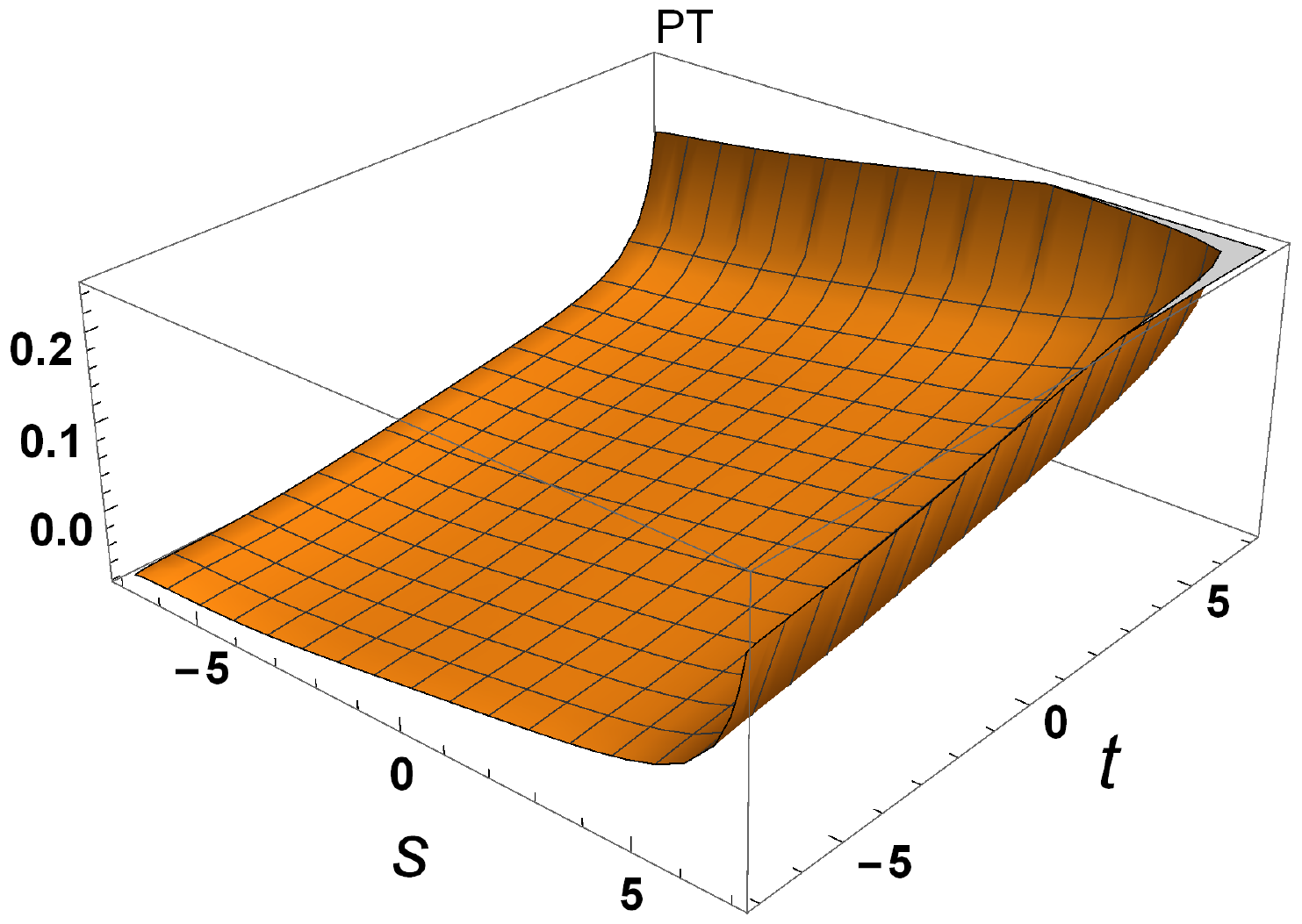}
\includegraphics[width=0.3\textwidth]{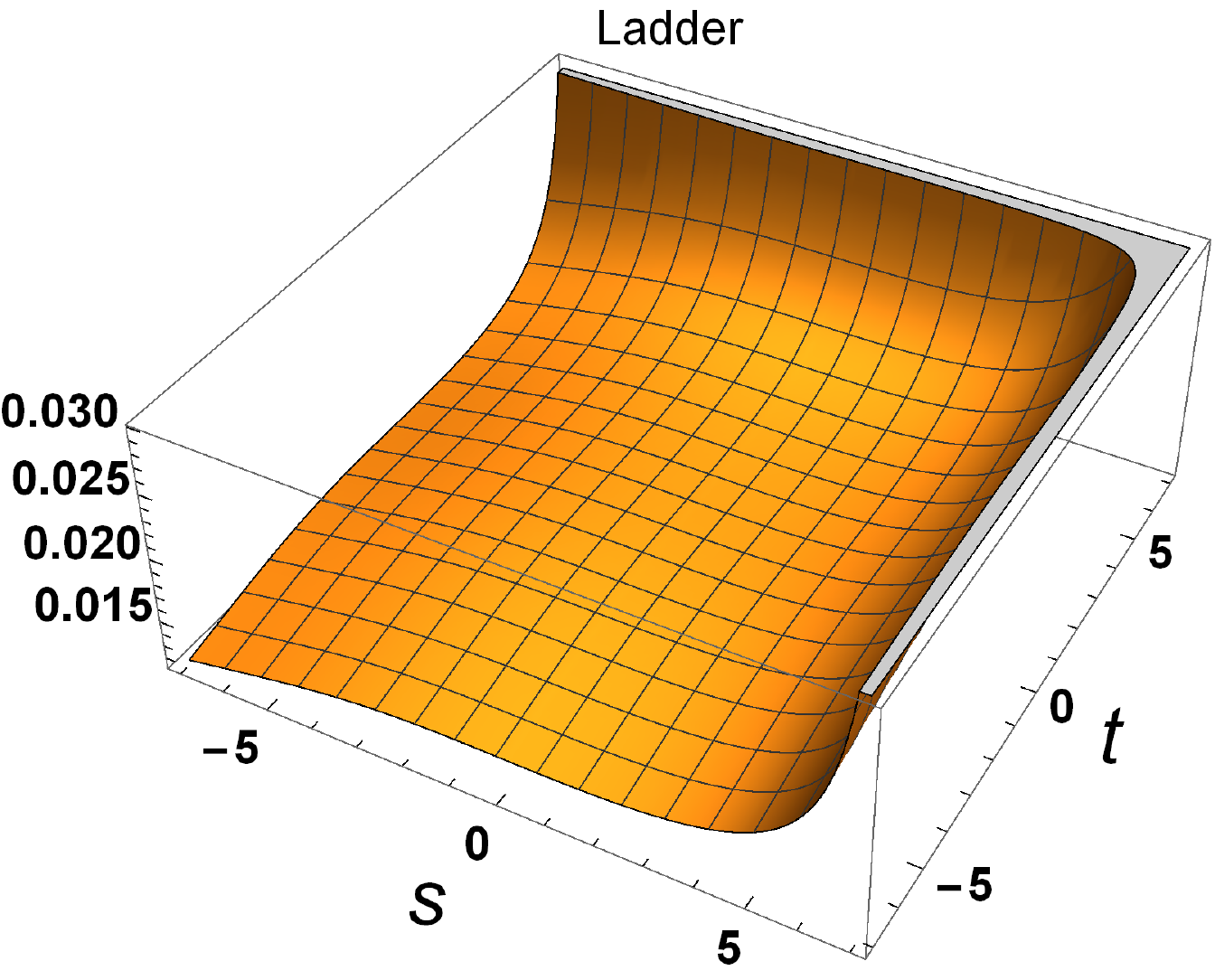}
\includegraphics[width=0.3\textwidth]{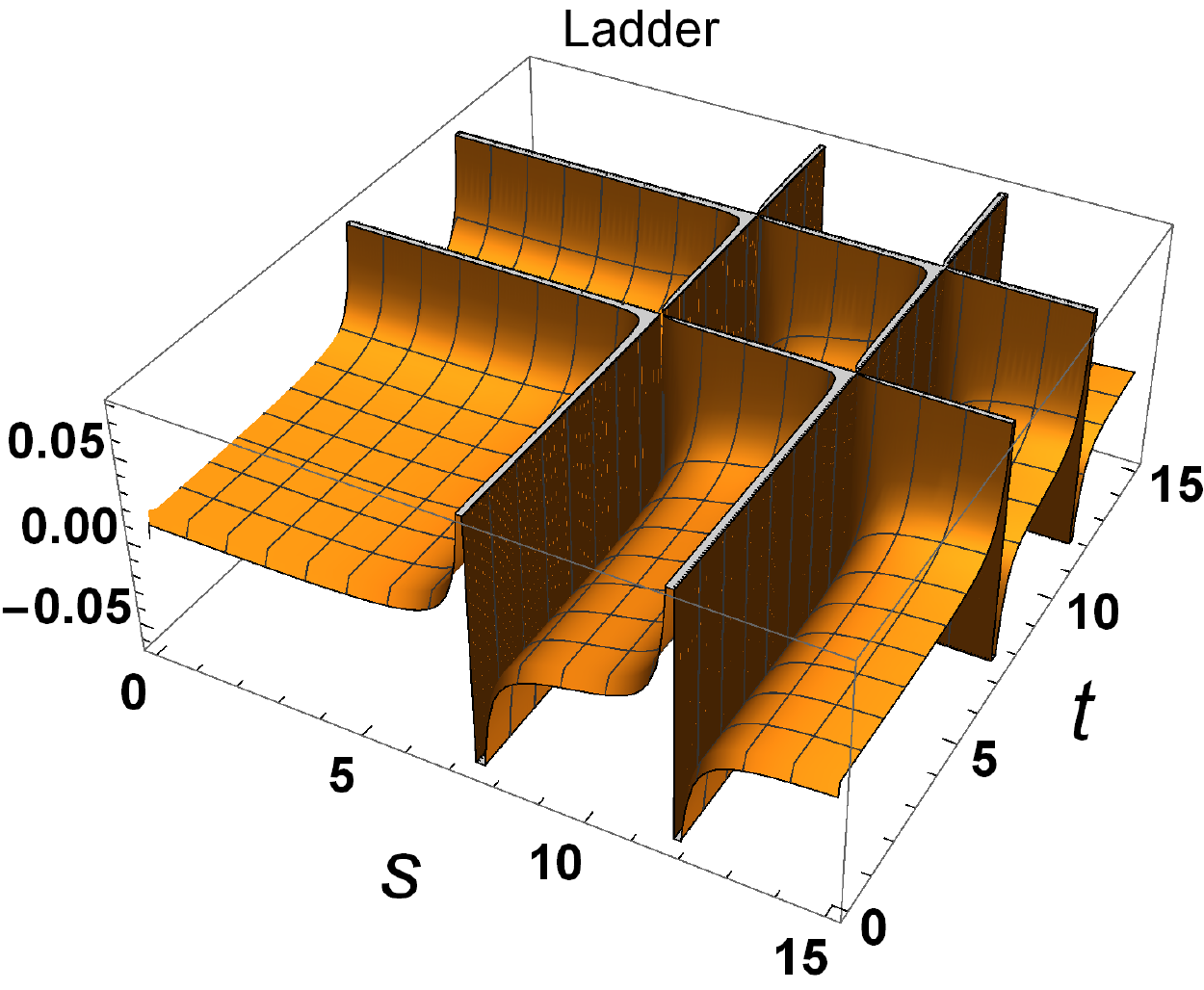}
\end{center}
\hspace{3cm} a \hspace{4cm} b \hspace{5cm} c
\caption{Comparison of PT(a)  and the  ladder approximation (b) in the region up to the first pole. The last plot (c) shows the ladder approximation beyond the first pole. One can clearly see the pole structure of the function $\Sigma$. }
\label{3d10}
\end{figure}

The comparison of all the curves is shown in Fig.\ref{allloop10}.
In Fig.\ref{3d10}, we also show the 3-dimensional plot.
The situation here is the same as in the case of $D=8$. The ladder approximation works pretty well and its analytical solution qualitatively describes all the features of the full equation. The function $\Sigma$ possesses an infinite number of periodic poles and one separate pole comes from  $\Delta$ (\ref{ladd}). There is no UV finite limit.

\section{Subleading divergences in D=8}
The analysis of the leading divergences performed in the previous sections can be extended to the subleading ones.
Despite the non-renormalizability of the theory due to locality of the $\R'$-operation the leading divergences in all loops are governed by the one loop term while the subleading divergences by the two loop ones. We showed how this procedure  explicitly  works in the theories of interest in  \cite{we3}.  There the recursive relations for the subleading terms as well as the differential equations for the all loop sum were obtained for the ladder type diagrams in D=8.
In principle, one can get the corresponding relations for the full set of diagrams; however, on the one hand, they are too cumbersome and, on the other hand, the ladder type diagrams seem to give a very good approximation to the full result and allow for the analytic solution.

\begin{figure}[htb!]
\begin{center}
\leavevmode
\includegraphics[width=0.45\textwidth]{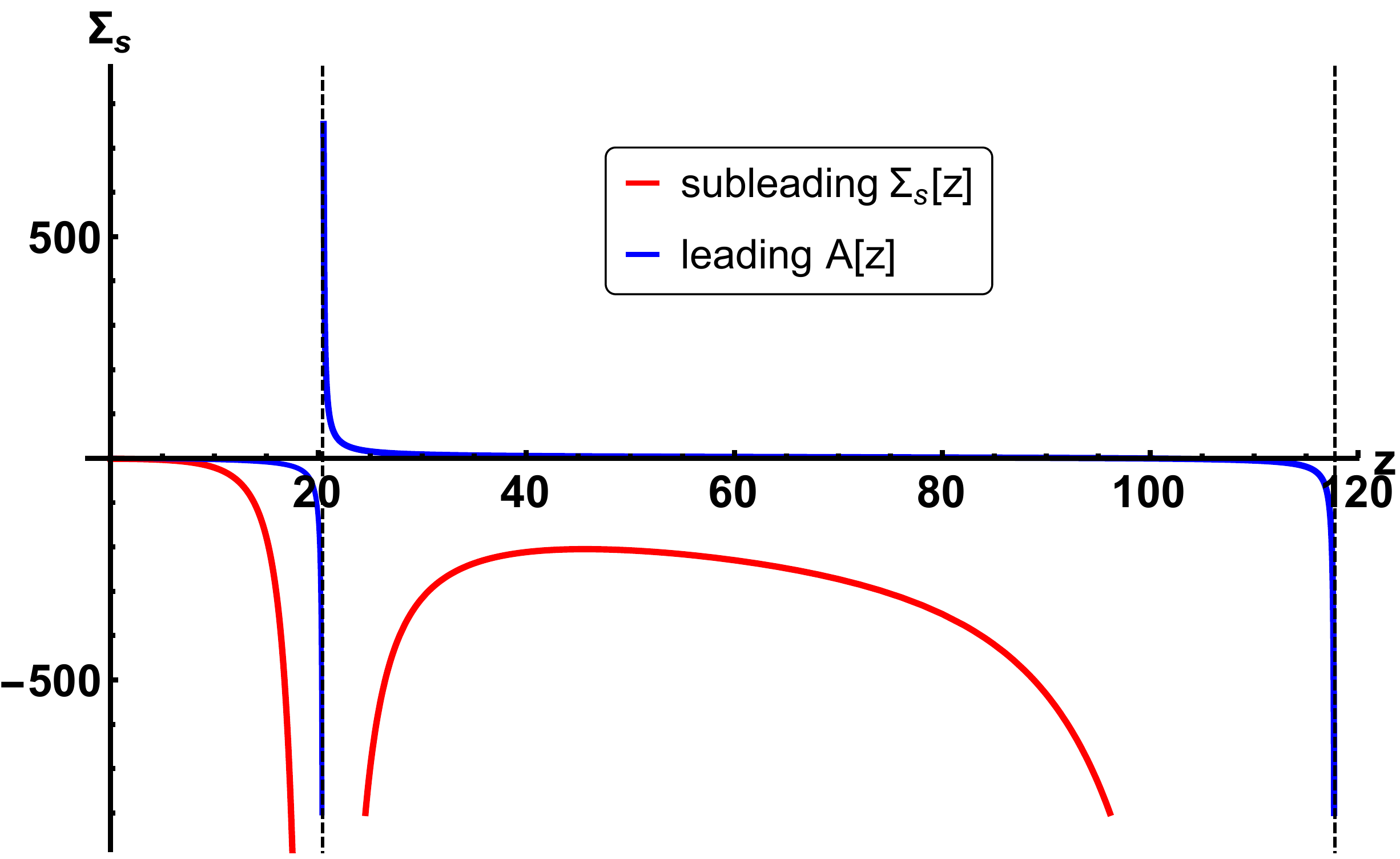}
\includegraphics[width=0.45\textwidth]{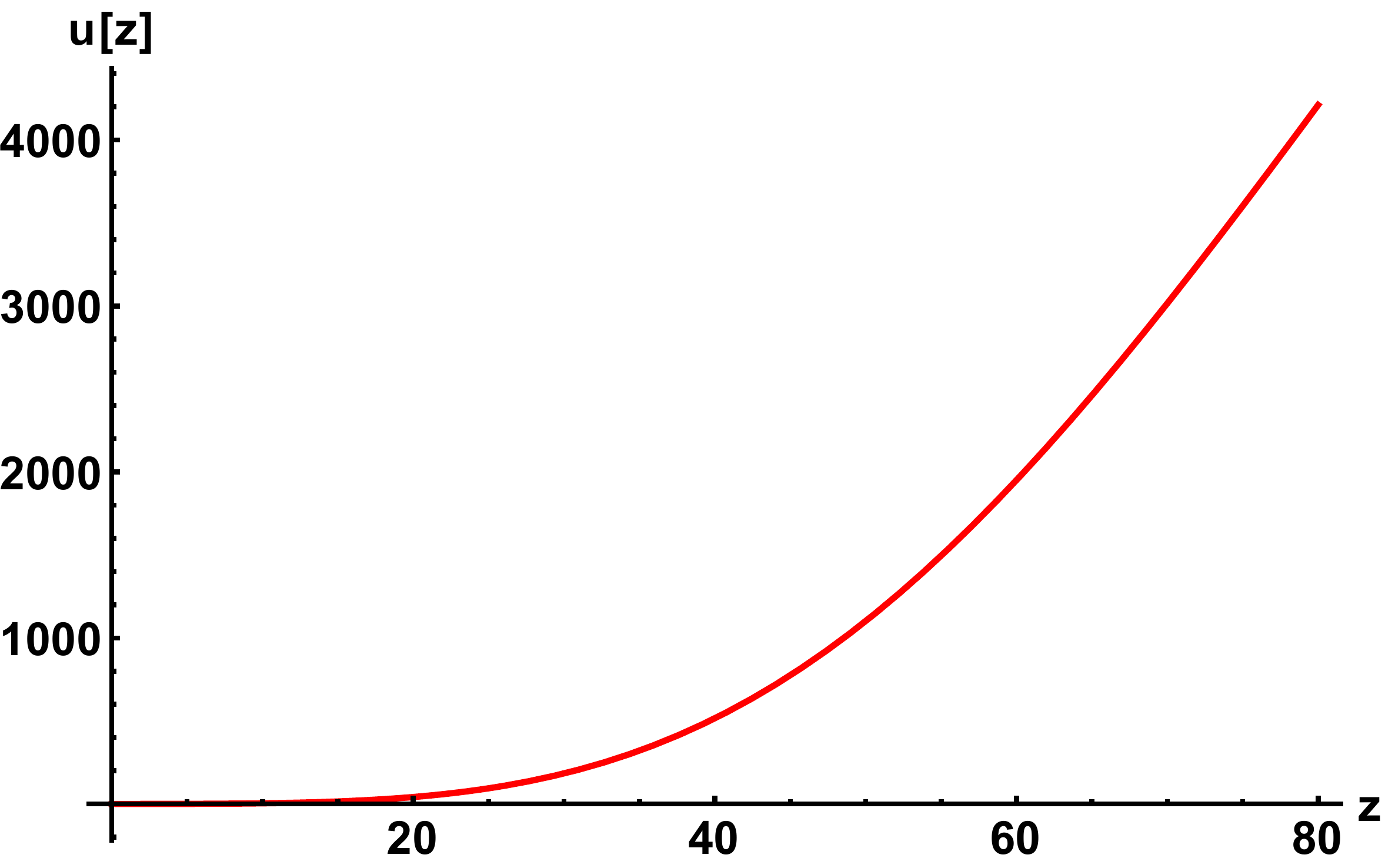}
\end{center}
\caption{Numerical solution for $\Sigma_{s}$ together with $\Sigma_{L}$ (left). Numerical evaluation of u(x) (right)}
\label{sub}
\end{figure}

First of all let us remind the relations obtained in  \cite{we3}.  The  subleading divergences  for the s-ladder type diagrams in all loops  are given by the sum of two functions proportional to $s$ and $t$
\beq
\Sigma_{Sub}(s,t,z)=s\Sigma_{s}(s,z)+t\Sigma_{t}(s,z).
\eeq
These functions depend on the single  dimensionless argument $x=zs^2$ and are expressed in terms of some auxiliary functions $\Sigma'_{s}$ and $\Sigma'_{t}$ which obey the second order differential equations.
One has
\beqa
\frac{d^2 \Sigma'_{t}(x)}{dx^2}-\frac{1}{30}\frac{d \Sigma'_{t}(x)}{dx}+\frac{\Sigma'_{t}(x)}{720}=-\frac{1}{432} ,\\ \nonumber
\frac{d\Sigma_{t}(x)}{dx}=\frac{1}{60}x\frac{d \Sigma'_{t}(x)}{dx}-\frac{\Sigma'_{t}(x)}{60}-x\frac{\Sigma'_{t}(x)}{720}-\frac{x}{432}.\label{sub_eq_t}
\eeqa
with the solution
\beqa
\Sigma'_{t}(x)=\frac{5}{6}\left[e^{x/60}(-sin[x/30]+2cos[x/30])-2\right],\\ \nonumber
\Sigma_{t}(x)=-\frac{1}{36}\left[60 + x + e^{x/60} (-(60 + x) cos[x/30] - 2 (-15 + x) sin[x/30])\right]. \label{sub_sol_t}
\eeqa
For the function $\Sigma'_{s}(x)$ the equation is slightly more complicated
\beq
\frac{d^2 \Sigma'_{s}(x)}{dx^2}+f_1(x)\frac{d \Sigma'_{s}(x)}{dx}+f_2(x)\Sigma'_{s}(x)=f_3(x), \label{sub_eq_s}
\eeq
where
\beqa
f_1(x)&=&-\frac 16+\frac{\Sigma_L}{15},\nonumber\\
f_2(x)&=&\frac{1}{80}-\frac{\Sigma_L}{120}+\frac{\Sigma_L^2}{600}+\frac{1}{15}\frac{d \Sigma_L}{dx}, \nonumber\\
f_3(x)&=&\frac{2321}{5!5!2}\Sigma_L+\frac{11}{1800}\Sigma'_{t}-\frac{469}{5!90}\Sigma_L^2-\frac{442}{5!5!6}\Sigma_L\Sigma'_{t}+\frac{23}{6750}\Sigma_L^3+\frac{1}{1200}\Sigma_L^2\Sigma'_{t}\nonumber\\
&-&\frac{19}{36}\frac{d \Sigma_L}{dx}-\frac{1}{15}\frac{d \Sigma'_{t}}{dx}
+\frac{23}{225}\frac{d \Sigma_L^2}{dx}+\frac{1}{30}\frac{d( \Sigma_L\Sigma'_{t})}{dx}-\frac{3}{32}.\nonumber
\eeqa
 $\Sigma_L(x)$ being the leading ladder  given by eq.(\ref{lad8}).
Then the function $\Sigma_{s}(x)$ is given by
\beq
\Sigma_{s}(x)=(x\frac{d}{dx}-1)\Sigma'_{s}-x(-\frac{19}{72}\Sigma_L+\frac{1}{12}\Sigma'_{s}-\frac{1}{30}\Sigma'_{t}+\frac{23}{450}\Sigma_L^2-\frac{1}{30}\Sigma_L \Sigma'_{s}+\frac{1}{60}\Sigma_L \Sigma'_{t}) .\label{Ric2}
\eeq
The above equation for $\Sigma'_{s}$ is difficult to solve and analyze analytically. However, from the form of the r.h.s one can deduce that it has poles governed by $\Sigma_L$.  Indeed, numerical study of this equation proves that
$\Sigma'_{s}$ (and $\Sigma_{s}$) inherits all the poles of $\Sigma_{L}$, as can be seen in Fig.\ref{sub} on the left where we have plotted the numerical solution for  $\Sigma_{s}$ together with  $\Sigma_{L}$.

Further, the linear inhomogeneous differential equation  (\ref{sub_eq_s}) can be simplified by the substitution  $\Sigma'_{s}(x)=\frac{d\Sigma_L}{dx} u(x)$ so that the function $u(x)$ obeys the equation~\cite{we3}
\beq
u''(x)= f_3(x)/\frac{d\Sigma_L}{dx}.\label{equ}
\eeq
Numerical integration of this equation reveals that the function $u(x)$ is a regular function without singularities (see Fig.\ref{sub} on the right). Hence the functions $\Sigma'_{s}(x)$ and $\Sigma_{s}(x)$ are essentially governed by $\frac{d\Sigma_L}{dx}$ and follow its analytical properties.

Thus, we conclude that the subleading divergences follow the pattern of the leading ones and have the same properties.
The sum of the subleading contributions in all loops behaves like the sum of the leading poles and has no finite UV limit.

\section{Discussion}

Summarizing the presented analysis of the UV divergences in maximally supersymmetric gauge theories  we have to underline once more that the leading as well as the subleading divergences are governed by the structure of the $\R'$-operation and can be evaluated starting from the one and two loop diagrams, respectively, with the help of the RG equations. This reasoning is valid for all further subsubleading divergent terms.

We have presented explicit formulas confirming this statement and have demonstrated how the higher terms can be calculated from the lower ones via pure algebraic recursive relations. The corresponding equations generalize the usual
renormalization group relations for the pole terms in the case of non-renormalizable interactions.These equations take the integro-differential form and do not admit simple solutions. They can be simplified for particular sets of diagrams, as can be seen by the example of the ladder type graphs.

Even this task happened to be quite complicated when going to the subleading terms and we were bound to use numerical methods.  However, the result of summation of subleading divergences does not lead to any qualitative difference from the leading terms. All the main features of the leading divergences keep untouched.

The numerical solution of these RG equations shows that the leading divergences qualitatively are very well described by the ladder approximation which admits the explicit analytical solution and can be analysed.  The form of this solution in $D=6,8$ and $10$ dimensions suggests that  one cannot get rid of the UV divergences by simply removing the regularization since the obtained solutions have no limit when $\epsilon\to 0$.  Indeed, in $D=6$  this limit depends on the kinematics and there is no way to  make all the amplitudes finite in all channels simultaneously. As for the $D=8$ and $10$ case, the corresponding functions have an infinite number of poles and the point $\epsilon=0$ is unreachable.

Thus, we see that the maximally supersymmetric gauge theories, despite numerous cancelations of divergent diagrams (the bubbles and triangles),
still contain the  divergent diagrams in all orders of PT even on-shell and their summation to all orders does not improve the situation.
This  means that these theories remain non-renormalizable and the knowledge of the all order result does not help to get a finite theory. It might be different if the summation procedure produced the function which possesses the finite limit in removing the regularization, which was the original motivation of this analysis.

This conclusion might seem disappointing; however, we believe that the knowledge of the infinite series of divergent terms can be useful to remove the arbitrariness of the subtraction procedure and get a meaningful theory. This theory, however,  would contain an infinite number of terms in the Lagrangian. Since these theories can be considered as the world-volume theories of stacks of branes it is interesting to look at problem from this end. However, the properties of the UV-completion in this case remain to be explored.

\section*{Acknowledgements}
 The authors are grateful to M.Kompaniets for numerical check of the 3- and 4-loop ladder type diagrams and to L.Bork for numerous useful discussions. This work was supported by the Russian Science Foundation grant \# 16-12-10306.

\end{document}